Solar irradiance changes and phytoplankton productivity in Earth's ocean following

astrophysical ionizing radiation events

Running title: Ionizing radiation events and marine productivity

Patrick J. Neale[1*] and  Brian C. Thomas[2]

[1]Smithsonian Environmental Research Center, Edgewater, MD

[2]Washburn University, Department of Physics and Astronomy, Topeka, KS;

[*]Corresponding author: P.O. Box 28, Edgewater MD, 21037, USA

nealep@si.edu

Tel:  443-482-2285

Fax: 443-482-2380

Abstract

Two atmospheric responses to simulated astrophysical ionizing radiation events significant to life on Earth are production of odd-nitrogen species, especially $NO_2$, and subsequent depletion of stratospheric ozone. Ozone depletion increases incident short-wavelength ultraviolet radiation (UVB, 280-315 nm) and longer ( > 600 nm) wavelengths of photosynthetically available radiation (PAR, 400 -700 nm). On the other hand, the $NO_2$ haze decreases atmospheric transmission in the long-wavelength UVA (315-400 nm) and short wavelength PAR. Here we use the results of previous simulations of incident spectral irradiance following an ionizing radiation event to predict changes in Terran productivity focusing on photosynthesis of marine phytoplankton. The prediction is based on a spectral model of photosynthetic response developed for the dominant genera in central regions of the ocean (*Synechococcus* and *Prochlorococcus*), and remote-sensing based observations of spectral water transparency, temperature, wind speed and mixed layer depth. Predicted productivity declined after a simulated ionizing event, but the effect integrated over the water column was small. For integrations taking into account the full depth range of PAR transmission (down to 0.1% of utilizable PAR), the decrease was at most 2-3% (depending on strain), with larger effects (5-7%) for integrations just to the depth of the surface mixed layer. The deeper integrations were most affected by the decreased utilizable PAR at depth due to the $NO_2$ haze, whereas shallower integrations were most affected by the increased surface UV. Several factors tended to dampen the magnitude of productivity responses relative to increases in surface damaging radiation, e.g. most inhibition in the modeled strains is caused by UVA and PAR, and the greatest relative increase in damaging exposure is predicted to occur in the winter when UV and productivity are low.

1. Introduction:

Astrophysical ionizing radiation can threaten life on Earth in several ways, but studies have indicated that the most significant effect is atmospheric ionization, causing catalytic destruction of ozone and consequent increase in surface UVB (280-315 nm; Atri et al., 2014; Ejzak et al., 2007; Gehrels et al., 2003; Reid et al., 1978; Thomas, 2009; Thomas et al., 2005). A variety of events are important, including gamma-ray bursts, supernovae (SNe), and solar proton events (SPEs). Depending on the event's energy fluence and spectrum, severe stratospheric ozone depletion can result, lasting for several years (Ejzak et al., 2007; Thomas et al., 2005). Such events also result in a "brown haze" associated with the increase in $NO_2$, which decreases incident irradiance in the longer-wavelength UVA (315-400 nm) and visible or photosynthetically available radiation (PAR, 400-700 nm) range. This phenomenon tends to dissipate more rapidly than the ozone depletion.

Phytoplankton photosynthesis in the upper layer of the ocean accounts for most marine productivity and constitutes about half of the plant productivity on Earth. Photosynthetic performance is dependent on the penetration of PAR into the ocean water column, but is negatively affected by exposure to UV in near surface waters.

Existing levels of UV are sufficient to cause widespread near-surface photoinhibition in most natural waters at least episodically (Harrison and Smith, 2009; Villafañe et al., 2003).  Studies of the spectral dependence of UV effects show that most inhibition is caused by UVA with UVB making a secondary contribution (Neale, 2000).  Severe ozone depletion over Antarctica as typically occurs in association with the springtime polar vortex has been estimated to possibly decrease productivity by 5-12% (Smith and Cullen, 1995; Smith et al., 1992) but studies making such estimates are mostly confined to the regions of the Southern Ocean and coastal Antarctic waters.

Another effect of ionizing radiation events that could reduce phytoplankton productivity is the decrease in incident PAR associated with the $NO_2$ haze.  Spectral dependence also needs to be considered in this response.  $NO_2$ haze particularly affects the UVA-blue and blue-green parts of the spectrum (350-500 nm) (discussed in more detail in Thomas et al. 2015).  Blue to blue-green irradiance is critical to marine productivity as it is the least absorbed by seawater and thus the deepest penetrating into the ocean.  The light harvesting complexes of most phytoplankton have strong absorption peaks in this spectral region (Larkum, 2003).  Accurate estimates of phytoplankton photosynthesis at depth thus depend on fully characterizing the profile of spectral irradiance and the proportion of that irradiance that is absorbed by phytoplankton photosynthetic complexes.

We have previously used several weighting functions to estimate impacts as direct changes in damaging irradiance at Earth's surface (Thomas et al., 2015). To gain a fuller understanding of how phytoplankton productivity would be affected by an ionizing radiation event, models are needed that take into account responses over the full spectral range. The models should take into account transmission of light into the water column, the response of photosynthesis to different wavelengths of UV and PAR and integrate effects on primary productivity over depth. Transmission of UV and PAR into the present day ocean can be estimated from remote sensing data using a variety of approaches (Fichot et al., 2008; Lee et al., 2013). Phytoplankton photosynthetic response to full spectral underwater irradiance has been modeled using a combination of biological weighting functions for the UV and photosynthesis-irradiance curves in the PAR, the so-called BWF/P-E model (Cullen et al., 1992; Neale, 2000). Cullen et al. (2012) defined efficient numerical approximations to BWF/P-E models for a range of phytoplankton species in order to facilitate calculations on a global basis. However, Cullen et al. (2012) did not have available for their analysis BWF/P-E models appropriate for the phytoplankton most prevalent in mid-ocean waters, (pico)phytoplankton, in the genera *Synechococcus* and *Prochlorococcus*. More recently, Neale et al. (2014) developed a new BWF/P-E model ($E_{max}$ model) that accounts for distinctive features of picophytoplanktonic photosynthetic response to PAR and UV radiation. Here, we report on the application of this model to productivity assessments over broad areas of the Earth's oceans. Effects have been evaluated over selected latitude and longitude ranges in the Pacific Ocean where we estimated spectral water transparency from satellite observations of ocean color.

Simulations of atmospheric effects (primarily production of odd-nitrogen compounds and subsequent depletion of ozone) have been previously completed for a variety of astrophysical ionizing radiation events (Ejzak et al., 2007; Gehrels et al., 2003; Melott et al., 2005; Thomas et al., 2007; Thomas et al., 2013; Thomas et al., 2008; Thomas et al., 2005). Melott and Thomas (Melott and Thomas, 2011) reviewed a wide range of events and concluded that supernovae, short duration hard spectrum gamma-ray bursts (SHGRBs), and long duration soft spectrum gamma-ray bursts (LSGRBs) (respectively) are the most likely astrophysical events to pose serious threats to life on Earth over a few 100 million year time scales. Piran and Jimenez (2014) arrive at similar conclusions for LSGRBs. While the exact frequency and intensity is not yet known, recent evidence indicates that extreme solar events may be more damaging and more frequent than previously thought (Melott and Thomas, 2012; Thomas et al., 2013). Ejzak et al. (2007) showed that atmospheric changes depend on total energy and spectral hardness, not duration of the event. In this report we present estimates of the effect on marine primary productivity by a gamma-ray burst (GRB) case described in Thomas et al. (2015). These results will be broadly applicable to astrophysical ionizing radiation events of any type (e.g. supernovae) with similar energy fluence and spectrum. The methods developed here may be applied to any ionizing radiation event for which atmospheric conditions are known or have been modeled.

2. Methods:

2.1 *Modeling of photobiological impacts*

2.1. Overview of modeling:  Results of the atmospheric chemistry and radiative transfer modeling previously described by Thomas et al. (2015) were used as input to a model that computes the transmission of light through the water surface and into a column with specified optical properties.  The model then calculated photosynthesis rates for specific phytoplankton strains, resulting in photosynthesis rates at a range of depths in the water column, as well as a depth-integrated value. The calculation computed potential photosynthesis rates, and took into account inhibition by both UV and visible light.  Since the atmospheric chemistry was only 2-D (no longitude) and the model is overall numerically intensive, the full simulation covered only selected regions of the ocean.  These correspond to the longitude range between 160°W and 140°W, at all latitudes, which is an area of the mid-Pacific Basin near the longitude of the Hawaiian Islands; and a longitude range between 110°W and 90°W, between 15°S and 15°N in latitude, a region near the longitude of the Galapagos Islands.  These regions were chosen to be representative of the overall variation in exposure and phytoplankton biomass in Earth's ocean (basis for selection is presented in section §2.3). All other longitude-dependent data were then processed using the same regions.  The following describes each of these steps in detail.

## 2.2 Spectral irradiance in surface water

As described by Thomas et al. (2005), the NASA Goddard Space Flight Center two-dimensional (latitude, altitude) time-dependent atmospheric chemistry and dynamics model (hereafter referred to as the "GSFC model") was used to compute the concentration and distribution of ozone, $NO_2$ and other optically important gases.  The GSFC model produced atmospheric profiles for 10° latitude bands centered on 85° N, 75° N, etc, for both a normal and GRB affected atmospheres (see details in Thomas et al. 2005).  Direct and diffuse spectral irradiance (over the range 280 – 700 nm) at the surface, mid-day, was then computed by the Tropospheric Ultraviolet and Visible (TUV) radiative transfer model (Madronich and Flocke, 1997), for each latitude band and daily for a five year period following the GRB (arbitrarily chosen as June 2000 to 2005) as described by Thomas et al. (2015).  The irradiance components were transferred through the ocean surface following the procedure outlined in Arrigo et al. (2003).  In brief, surface reflectance was calculated for direct and diffuse irradiance, with each component being a sum of specular reflection and reflection from sea foam.  Direct and diffuse irradiance was then weighted by these reflectance coefficients and summed to yield total spectral irradiance under the surface ($E(0^-,\lambda)$, W $m^{-2}$ $nm^{-1}$).  Reflection from sea foam was calculated using relations from Gregg and Carder (Gregg and Carder, 1990), which depend on wind speed.  For a flat ocean, specular reflection was found using Fresnel's law, which depends on the solar zenith angle and the refracted angle in water.  For wind speeds greater than 2 m s$^{-1}$, we used an empirical relation from Gregg and Carder (1990) and wind speeds from the The Scatterometer Climatology of Ocean Winds (SCOW, http://cioss.coas.oregonstate.edu/scow/ ) (Risien and Chelton, 2008). We used monthly wind speed maps, with values averaged over each longitude range and 10 degree latitude bins to match the GSFC model output.

2.3 Spectral attenuation of irradiance:

The penetration of irradiance through the ocean's water column was calculated from spectral attenuation coefficients ($K_d(\lambda)$, m$^{-1}$) which were estimated for the chosen ocean regions based on climatological satellite remote sensing data.  We used $K_d(\lambda)$ values that are time and location dependent.  The values are particularly important in the UV part of the spectrum, since this is where the main inhibition of photosynthesis occurs, and the water clarity strongly affects the penetration of these wavelengths.  Fichot and Miller (2010) have used the SeaUV algorithm (Fichot et al., 2008) to compute monthly climatologies of $K_d(\lambda)$ at $\lambda$ = 320, 340, 380, 412, 443, and 490 nm based on SeaWiFS remotely-sensed reflectance data.  We used their global climatology of $K_d$ to select exposure regions representative of global variation (§2.3 and Neale et al. unpublished). Once the exposure regions were selected, we updated the $K_d$ values by applying the Fichot et al. (2008) methodology to the 2010 reprocessed SeaWiFS climatology which was also used to calculate visible range $K_d$'s (412-700 nm, 1 nm resolution) based on the Quasi-Analytical Algorithm (QAA) v6 (Lee et al., 2013). $K_d$ values over the full range of 280-412 nm at 1 nm resolution were obtained from the SeaUV estimates by interpolation/ extrapolation of log transformed values (i.e. assuming an approximate exponential dependence of $K_d$ on wavelength in the UV). Based on the recommendations of Lee et al. (2013), the SeaUV $K_d$'s were adjusted to agree with QAA v6 output at 412 nm, with the magnitude of the offset proportionally reduced with decreasing wavelengths to 0 at 320 nm. All latitude-dependent data were averaged over 10 degree latitude bands for each time point to match the output of the GSFC model. More detail on the estimation of the $K_d (\lambda)$ will be reported elsewhere.

2.3  Transparency depths

We defined two metrics for inhibiting and photosynthetically utilizable irradiance, $T_{pir}$ and $T_{pur}$, respectively (Lehmann et al., 2004). These were used to gauge the overall effect of a GRB on the exposure of oceanic plankton to solar radiation taking into account water column transparency. Transparency to inhibiting irradiance ($T_{PIR}$) is defined as:

$$T_{PIR} = \sum_{280nm}^{700nm} \frac{1}{K_d(\lambda)} \varepsilon(\lambda) E(0^-, \lambda) \Delta\lambda \qquad (1)$$

where $\varepsilon(\lambda)$ (m$^2$ W$^{-1}$) are the biological weighting function (BWF) coefficients for UV inhibition of photosynthesis (Neale, 2000) and E(0$^-$,$\lambda$) (W m$^{-2}$ nm$^{-1}$) is spectral irradiance just below the surface (see §2.2) . The selection of BWFs for the calculation is discussed in §2.5. A constant, average $\varepsilon(\lambda)$ over the 400-700 nm range, $\varepsilon_{PAR,}$ was used to represent inhibition by PAR. The transparency to photosynthetically utilizable radiation ($T_{PUR}$) is defined as

$$T_{PUR} = \sum_{400nm}^{700nm} \frac{1}{K_d(\lambda)} \frac{a_p(\lambda)}{\overline{a_p}} \frac{E(0^-,\lambda)}{E_{PAR}(0^-)} \Delta\lambda \qquad (2)$$

where $a_p(\lambda)$, (m² mg Chl⁻¹) are the chlorophyll-specific spectral absorption coefficients for each species used in the simulation (more details in section 2.5), $\overline{a_p}$ is average $a_p(\lambda)$ and $E_{PAR}(0^-)$ is PAR just below the surface (W m⁻²). Representative longitudinal ranges were chosen based on the geographic distribution of the annual mean $T_{PIR}/T_{PUR}$ ratio calculated from the monthly $K_d$ climatology (see §2.2). Figure 1 shows the variation in zonal average $T_{PIR}/T_{PUR}$ ratio (±SD) for all ocean basins (global) compared to the averages for our two chosen longitudinal ranges, mid-Pacific (140°-160° W) and East-Pacific (90°-110° W). In general, we chose the Pacific basin as having the greatest continuous latitudinal range of open ocean. Within the Pacific, we chose the mid-Pacific as the best overall match for the global average (Fig. 1). However, the mid-Pacific is higher than the global average in two bands north and south of the equator at 15° S to 15 N° (Fig. 1). Thus we chose a second region in the East-Pacific that matches global average well at those latitudes. These averages use the BWF and $a_p$ of just for one species (Syn1 as described in §2.5). Similar results were obtained using parameters for other species. A more comprehensive exposition on the global distribution of $T_{PIR}/T_{PUR}$ will be provided in a separate report.

## 2.4 Depth Integrated Photosynthesis

Photosynthesis rates at each depth point, $z$, were computed following a procedure similar to that described in Lehmann et al. (2004), with some modifications to enable use with the BWF photosynthesis-irradiance (BWF/P-E) $E_{max}$ model as described by Neale et al. (2014). This model was originally formulated to represent photosynthetic response in a laboratory incubator (photoinhibitron) using filtered irradiance from a xenon arc lamp. In order to apply the model to water column conditions, a correction factor was applied to account for the spectral differences between incubator PAR and underwater PAR at each depth. Irradiance weighted chlorophyll-specific absorption for the photoinhibitron, $a_{PI}$ (m$^2$ mg Chl$^{-1}$), was calculated by weighting $a_p(\lambda)$ with the average photoinhibitron spectrum, as photon flux, $E_{PI}^{Q}(\lambda)$,

$$a_{PI} = \sum_{400\,nm}^{700\,nm} a_p(\lambda) \cdot E_{PI}^{Q}(\lambda) \cdot \Delta\lambda \left/ \sum_{400\,nm}^{700\,nm} E_{PI}^{Q}(\lambda) \cdot \Delta\lambda \right.$$

$$(3)$$

The calculation was based on photon (quantum) flux ($E^{Q}(\lambda)$, μmol photons m$^{-2}$ s$^{-1}$ nm$^{-1}$) since photosynthesis is a quantum process. The wavelength resolution ($\Delta\lambda$) was 1 nm. A similar calculation was performed for the underwater profile to obtain the irradiance weighted absorption of in situ irradiance, $a_{IS}(z)$ (m$^2$ mg Chl$^{-1}$):

$$a_{IS}(z) = \sum_{400\,nm}^{700\,nm} a_p(\lambda) \cdot E^{Q}(0^-,\lambda) \cdot e^{-K_d(\lambda)z} \cdot \Delta\lambda \left/ \sum_{400\,nm}^{700\,nm} E^{Q}(0^-,\lambda) \cdot e^{-K_d(\lambda)z} \cdot \Delta\lambda \right.$$

$$(4)$$

Finally, a corrected PAR irradiance for the photosynthesis model was calculated as

$$E'_{PAR}(z) = E_{PAR}(z) \frac{a_{IS}}{a_{PI}}$$

(5)

For consistency with previous usage of the BWF/P-E model, the corrected PAR is in energy units (W m$^{-2}$). The correction adjusted $E_{PAR}$ for the better (or worse) quantum effectiveness of underwater irradiance compared to photoinhibitron irradiance; the factor was applied to both light-limited photosynthesis at low irradiance and PAR inhibition at high irradiance (pigments mediate both processes). Next, the dimensionless dose rate of photosynthesis-inhibiting radiation, $E_{PIR}$, was calculated using an appropriate biological weighting function, $\varepsilon(\lambda)$ and a weight for PAR inhibition, $\varepsilon_{PAR}$:

$$E_{PIR}(z) = \sum_{280\,nm}^{400\,nm} \varepsilon(\lambda) \cdot E(0^-,\lambda) \cdot e^{-K_d(\lambda)\cdot z} \cdot \Delta\lambda + \varepsilon_{PAR} \cdot E'_{PAR}(z)$$

(6)

These quantities were then used to calculate photosynthesis using the BWF/P-E model. The first part of the model expresses the potential rate of photosynthesis in the absence of inhibition ($P^B_{pot}$, g C g Chl$^{-1}$ h$^{-1}$):

$$P^B_{pot}(z) = P^B_s \cdot \left(1 - e^{-E'_{PAR}(z)/Es}\right)$$

(7)

where $P^B_s$ (g C g Chl$^{-1}$ h$^{-1}$) is the maximum rate of photosynthesis normalized to chlorophyll (Chl, g Chl m$^{-3}$) biomass, and $E_s$ (W m$^{-2}$) is the saturation parameter for photosynthesis. The realized rate of photosynthesis was estimated by adjusting the potential rate by a factor accounting for the effect of inhibition:

$$P^B(z) = P^B_{pot}(z) \cdot ERC(z) \tag{8}$$

where $ERC$ ($z$) (dimensionless) is the "Exposure Response Curve" that describes the relative reduction in photosynthesis as a function of inhibiting irradiance. Our calculations used the $E_{max}$ model as described by Neale et al. (2014) as most appropriate for describing the response of picophytoplankton to full spectral irradiance. The relative effect of inhibition follows a rectangular hyperbolic response at low exposure transitioning to an inverse response above a threshold $E_{max}$:

$$ERC(z) = \left\langle \begin{array}{ll} \dfrac{1}{(1+E^*_{inh})} & E^*_{inh} \leq E_{max} \\[2mm] \dfrac{1}{cE^*_{inh}} & E^*_{inh} > E_{max} \end{array} \right.$$

$$c = \frac{1 + E_{max}}{E_{max}} \tag{9}$$

Conceptually, $E_{max}$ defines the transition between the exposures for which repair increases with damage to higher exposures for which repair is constant (i.e. operating at some maximum rate). A depth integrated photosynthesis, normalized to biomass, ($P^B$z, g C m g Chl$^{-1}$) was calculated from $P^B(z)$ over two depth ranges: 1) the depth at which $E'_{PAR}$ falls to 0.1% of the surface $(0^-)$ $E'_{PAR}$; 2) the mixed layer depth.  The 0.1% LD can be considered a conservatively high estimate of productivity as it only considers the variation of light, which for many parts of the ocean extends deeper than the surface mixed layer.  Although many factors change below the mixed layer (most importantly a lower temperature), the model is still a good approximation of productivity.  Note that 0.1% LD for $E'_{PAR}$ covers the full PAR wavelength range (400-700 nm) taking into account pigment absorption and thus differs between strains.  On the other hand, the MLD estimate only considers the productivity above the thermocline; this is a lower estimate of productivity but the model is only being applied under conditions it was originally defined.

To estimate the effect of a GRB event, productivity estimates were compared for model calculations using radiative transfer irradiance under both normal and post-GRB conditions (Thomas et al., 2015).   A weighted average productivity was used as an overall metric of GRB effects.  It is the product $P^B_Z$, × surface Chl x day length (h) which (times a constant) is an approximate estimate of carbon production per day (g C m$^{-2}$ d$^{-1}$) (Cullen et al., 2012).  Day length for a given latitude and time of year was calculated by a Matlab function (day_length.m) and Chl is an ocean color climatology for the region (same data set used in 2.3).  The product was summed over latitude for each month for the GRB and non-GRB case and the ratio of the sums is the overall effect on productivity.

2.5 Selection of BWF/P-E parameters

The biological weighting functions and photosynthesis parameters were measured for cultures of *Synechococcus* and *Prochlorococcus* grown at two temperatures, 20° C and 26° C, and two $E^Q_{PAR}$ levels, ML (ca. 80 μmol photons m$^{-2}$ s$^{-1}$) and HL (ca. 200 μmol photons m$^{-2}$ s$^{-1}$) (Neale et al. 2014 and Neale et al. unpublished data). The cultures were obtained from the National Center for Marine Algae, *Synechococcus* strains CCMP1334 and CCMP2370 (herein labeled "Syn1" and "Syn2") and *Prochlorococcus* strain CCMP1986. Parameters used in the model calculations were chosen by reference to estimated water column temperature and average irradiance for each date and location. Sea surface temperatures were retrieved from the IRI/LDEO Climate Data Library (http://iridl.ldeo.columbia.edu). Specifically, we use the Reynolds and Smith (1995) global monthly climatologies. We again averaged over 10° latitude bands, and the longitude ranges described above and considered the surface temperature to apply to whole mixed layer.

Light saturated rate of photosynthesis ($P^B_s$) and light saturation parameter ($E_s$) are well known to vary with temperature according to an Arrhenius equation ($Q_{10}$) type response. Usually, $Q_{10}$ for phytoplankton photosynthetic parameters is around 2 (Eppley, 1972). Based on this assumption, temperature functions were estimated for $P^B_s$ and $E_s$ using an equation of the form $m_1 \exp(m_2 (T-20))$. $P^B_s$ and $E_s$ were not significantly different between the two growth irradiances (Neale et al. 2014), so all growth data for each temperature was pooled in estimating the equations (n=12-14). As expected, in most cases $m_2$ was not significantly different from 0.0693 (corresponding to a $Q_{10}$ of 2). To obtain weighted inhibiting irradiance ($E_{PIR}$) appropriate to the sea surface temperature between 20° and 26°C, $E_{PIR}$ was calculated for each temperature and linearly interpolated.  Fixed values were used for temperatures below 20° or above 26°.

To choose between BWFs obtained for the two growth irradiances, we first computed the average PAR irradiance in the upper ocean, defined by the mixed layer depth.  Monthly climatologies of mixed layer depth were retrieved from the Naval Research Laboratory website (http://www7320.nrlssc.navy.mil/nmld/nmld.html).  This global data was averaged over 10° latitude bands, and the longitude regions described above.  The average upper ocean PAR was then compared to the midpoint between the "high" (200 μmol photons m$^{-2}$ s$^{-1}$) and "low" (80 μmol photons m$^{-2}$ s$^{-1}$) irradiance BWF versions; for values less than 140 μmol photons m$^{-2}$ s$^{-1}$, the low BWF version was used, while for values greater than 140 μmol photons m$^{-2}$ s$^{-1}$, the high BWF version was used.

3. Results

We previously showed that GRBs cause an increase in incident irradiance weighted for inhibition of photosynthesis (Thomas et al., 2015).  In that work, we used modeling results for a GRB occurring over Earth's South Pole, in late June (roughly the Southern Hemisphere winter solstice).  This choice was motivated by several factors.  First, Melott and Thomas (2009) have argued that a South-Polar burst fits well with what is currently known about the late Ordovician mass-extinction (briefly, extinction rates were higher in the Southern hemisphere).  Second, we wished to examine an extreme (but still realistic) event, with severe $O_3$ depletion; a polar burst isolates the ozone-depleting compounds to that hemisphere, which tends to result in greater localized $O_3$ depletion.  Finally, the June case is a middle-ground example of maximum depletion for a polar burst (Thomas et al., 2005).

Here, we use $T_{PIR}$ as a measure of the exposure of phytoplankton in the surface layer of the ocean to inhibiting irradiance.  This metric takes into account both the intensity of UV at the surface and the transparency of the surface layer to UV (Eq. 1).  In addition, we use weighting functions for phytoplankton strains that are representative of genera most prevalent in the central regions of the Earth's oceans, two strains of *Synechococcus* and one strain of *Prochlorococcus*.  The distribution of $T_{PIR}$ over latitude and for a 5 year time period following a GRB is shown in Fig. 2.  A similar yearly cycle of exposure is shown for the response of all the examined strains, with $T_{PIR}$ peaking in each hemisphere's summer period and overall more inhibiting exposure in the southern hemisphere.  The latter is the result of the GRB effects being confined to the Southern hemisphere, as well as particularly clear conditions in the Southern mid-Pacific Ocean in the longitude range shown here (around the longitude of the Hawaiian Islands) (Tedetti et al., 2007).  Peak exposure in both hemispheres occurs around the latitude of 25°.  Even though incident irradiance is also high at the equinox at the Equator, inhibiting irradiance in the equatorial region (±5° of the equator) is lower because this is an area of lower water transparency due to upwelling and greater phytoplankton growth (e.g. Cullen et al., 2012 their Fig. 7a). This upwelling covers a wider latitude range in the E. Pacific equatorial region (around the Galapagos, ±15° of the equator) which shows even lower transparency than the mid-Pacific equatorial region (Fig. 2, lower panels) consistent with its lower $T_{PIR}/T_{PUR}$ ratio (Fig. 1). Over the five year period in the N. hemisphere, $T_{PIR}$ remains at the same level. In the S. hemisphere, the largest peak is seen in the first austral summer solstice following the GRB event (month 6), after which there is a gradual lowering of exposure over the five years (Fig. 4). The pattern corresponds to the time course of incident weighted irradiance in the S. hemisphere after the GRB event, driven by an initial reduction in stratospheric ozone (~50% at 25°S), followed by a gradual return close to pre-event levels over 5 years (Thomas et al., 2015). No significant changes occur in the N. hemisphere. Only small relative changes in $T_{PIR}$ over time (i.e. not evident at the scales shown in Fig. 2) were computed for the E. Pacific region, similar to the change in the same latitude range in the mid-Pacific around Hawaii (Fig. 2). Although the relative variation of $T_{PIR}$ is similar for all three studied strains, the magnitude of the $T_{PIR}$ differs between species. *Prochloroccus* has higher $T_{PIR}$ than *Synechococcus*, reflecting its overall greater sensitivity to inhibition.

We also assessed the effect of GRBs on how much photosynthetically utilizable radiation (PUR) is present in the ocean's surface layer using the metric $T_{PUR}$. Like $T_{PIR}$, $T_{PUR}$ takes into account both the incident irradiance and water transparency. Spectral irradiance between 400 and 700 nm is weighted by the relative pigment absorbance of each strain.  In this way, PUR differs from the more common irradiance metric, PAR (photosynthetically available irradiance), which weights all irradiance between 400-700 nm equally.  The five year time course of $T_{PUR}$ for region 1 shows a similar pattern as $T_{PIR}$, with summertime peaks around 25° latitude in each hemisphere, and overall more $T_{PUR}$ in the S. hemisphere (Fig. 3). $T_{PUR}$ in the equatorial region is also lower in the E. Pacific than mid-Pacific related to the more extensive upwelling zone in the E. Pacific.  The S. hemisphere peak is reduced right after the GRB event, relative to that at year 5 (detail in Fig. 4).  The reduction is caused by the presence of the $NO_2$ haze that develops immediately after the GRB event and absorbs strongly between 320 and 480 nm (Thomas et al., 2015).  This reduces $T_{PUR}$ because *Synechococcus* and *Prochlorococcus* pigments also absorb strongly in this range that moreover is the wavelength region of highest seawater transparency (Pope and Fry, 1997).  The reduction in $T_{PUR}$ is less noticeable after the $NO_2$ haze mostly dissipates in years three and following.  This is in contrast to changes in incident PAR which actually increases in years 3-5 (Thomas et al., 2015). These increases are due to the lower absorbance by ozone in the visible (the Chappuis bands).  However, irradiance is mainly increased in the wavelength range of 550-700 nm where water transparency and relative phytoplankton pigment absorption are both low, so there is little effect on $T_{PUR}$.

The overall effect of the GRB over the ocean water column is an increase in inhibiting irradiance and decrease in photosynthetically utilizable irradiance with the maximum effects in the first two years after which they gradually dissipate.  The greatest effect of the GRB is at 25°S where it increases $T_{PIR}$ for strain 1 of

*Synechococcus* by 1.3 m (max – min of annual running mean) and decreases $T_{PUR}$ by

0.6 m (cf. Fig. 4) .   In the N. Hemisphere, the change in annual mean $T_{PIR}$ and $T_{PUR}$ is

≤ 0.3 m and ≤ 0.1 m, respectively (data not shown). Given these results, the remaining report will focus on results for the S. Hemisphere.

To gauge the effect of these perturbations in the light environment on

phytoplankton primary production, we calculated the ratio of depth integrated

production, $P^B_Z$, between the GRB and non-GRB conditions.  First, we consider depth

integration to the 0.1% light depth (Fig. 5).  Predicted phytoplankton productivity

integrated over the full water column (0.1% LD) decreased 2-3% in the mid-latitude

southern hemisphere ocean following the GRB.  The effect is mainly seen in the first

two years after the event and the relative effect is most pronounced in the mid to

high southern latitudes in the fall of the first year after the GRB (about 10 months

after the event).  Interestingly, the relative response is greater for *Synechococcus*

than *Prochlorococcus*.  This might seem inconsistent with the greater sensitivity of

*Prochlorococcus* shown in Fig. 2.  In fact, it is a consequence of the non-linear relationship between inhibition and exposure (Eq. 9).  With a hyperbolic response, as the effect of a given absolute exposure increases, the relative effect of an additional increment in exposure decreases.

As mentioned above, the effect on $P^B_Z$, is due to both enhanced inhibition near the surface caused by increased UV as well as decreased photosynthesis at depth due to less PUR.  To separate out the contribution of these two effects, we performed a separate integration using the potential rate of photosynthesis, $P^B{}_{pot}(z)$ (Eq. 7). From this we obtained $P^B{}_{pot\text{-}z}$, depth integrated production based only on the potential rate which responds to changes in PUR but does not reflect inhibition. Again, we use the ratio of integrals for GRB and non-GRB conditions as a function of time and latitude (Fig. 6).  As can be seen by comparing Figs. 5 and 6, much of the effect on $P^B{}_z$, in the first year is accounted for by the effect on $P^B{}_{pot\text{-}z}$, i.e. because of a reduction in PUR.  The relative effect was greatest in the spring and summer when PAR is lower, and has a greater relative effect on productivity by *Prochlorococcus* which has more of its pigment absorption in the wavelength range affected by $NO_2$ absorption.

The effect of a GRB presents a different pattern when examined in terms of productivity integrated only to the depth of the mixed layer (MLD).  In this case, the greatest effects were seen for the summer around 25°S latitude (Fig. 7).  The reduction in $P^B{}_z$ for these conditions approaches 5-7% depending on strain.  Also, the greatest effects were seen in the second year following the GRB corresponding to the time of the greatest % change in column ozone (Thomas et al. 2015).  The dominance of the inhibition component is also seen by comparison with the GRB effect on potential productivity integrated to the MLD, $P^B{}_{pot\text{-}z,}$ which shows the effect of reduced PUR (Fig. 8).  There was almost no GRB effect in the SH summer, instead the effects were most apparent in spring and autumn (cf. Fig. 5 for the 0.1% LD integration).  Indeed, the % reduction due to lower PUR was about the same as for the 0.1%LD integration, i.e. 1-2% (colors are different because of the different scales).

To gauge the overall effect on the tropical-temperate S. hemisphere ocean productivity, we calculated the % change in biomass and daylength weighted productivity summed over all southern latitudes– a metric which controls for the relative contribution of phytoplankton at each latitude and time to hemispheric production. This was estimated for both $P^B_Z$ and $P^B_{pot-Z}$ (i.e. depth integrated production both including and excluding inhibition). The latter integral reflects just the effect of the GRB on PUR, as shown by the changes in $T_{PUR}$. A ratio of sums, GRB vs no-GRB, was calculated for each integral and the difference between the ratios (as a decrement from unity) is the "inhibition effect", the contribution of inhibition per se to the change in $P^B_Z$, vs the shading effect which is shown by the change in $P^B_{pot-Z}$. This calculation suggests that the overall reduction in integral productivity in the S. Hemisphere following a GRB would be in the range of 1-2% (Fig. 9). The shading effect on PUR is the dominant factor for integration to the 0.1% LD and is important even for the MLD. The shading effect is most evident in the first three years, after which the effect on $P^B_Z$ was much smaller and mostly due to inhibition. Inhibition was greatest in years 2 and 3, consistent with time courses of incident weighted irradiance (Thomas et al 2015). Surprisingly, in the first six months, especially for the *Prochlorococcus* case, relative inhibition was actually LESS in the GRB case than the no-GRB case, so we get an inhibition effect > 1 (effectively an enhancement). This is attributed to the reduction in UVA caused by the $NO_2$ which was strongest early on (Thomas et al. 2015). UVA exposure is an important, indeed dominant, contributor to photoinhibition particularly in picophytoplankton strains (Neale et al. 2014). The weighted irradiance at the surface was always higher for the GRB case than non-GRB (as shown in Thomas et al. 2015), but at depth, after the preferential loss of the shorter wavelengths enhanced by GRB, the reduction in UVA due to $NO_2$ becomes the most important effect. In most cases, however, it did not outweigh the shading effect so that average $P^B_Z$ was still lower for GRB conditions vs non-GRB but for Pro MLD case, overall average $P^B_Z$ was actually slightly higher for the GRB case early in the record (Fig. 9,f).

4.0 Discussion:

We have performed a detailed assessment of how water column productivity in the tropical and temperate areas of the ocean would be affected by a specific type of gamma-ray burst. These results may be applied to any astrophysical ionizing radiation event of similar energy fluence and spectral hardness, and the method applied to any ionizing radiation event for which atmospheric conditions are known or have been modeled. Generally, the effect on productivity was small, < 7% reduction for integration to the mixed layer depth (MLD) and < 2% for integration to the 0.1% light depth. The increase in surface UVB due to a GRB event did have a substantial effect at the surface, decreasing photosynthesis [$P^B(0)$] on average around 5-7 % (Neale et al. 2014) and as much as 19% (maximum effect for all simulated conditions). But even the inhibition effect at the surface seems less than might be expected given the results of Thomas et al. (2015) that showed several fold increase of incident irradiance weighted for inhibition.

Several factors tended minimize the effects of increased damaging irradiance on primary productivity based on the response of *Synechococcus* and *Prochlorococcus* including:

1) Compared to the BWFs used in Thomas et al. (2015) (cf. Neale 2000), BWFs for *Synechococcus* and *Prochlorococcus* have a much greater sensitivity to UVA and PAR (Neale et al. 2014). This, to a certain extent, buffered the effect of additional UVB, i.e. the relative increase in weighted irradiance was small because it was added to the already substantial weighted exposure from UVA and PAR. Indeed, early on in the series, UVA exposure was actually less after a GRB because of heightened $NO_2$ absorbance (Thomas et al. 2015, their Fig. 12). Note also that the calculations of weighted irradiance in Thomas et al. (2015) did not include PAR contributions (though these are small for the functions that were used).

2) In the central ocean, penetration of UVB is shallower than UVA and PUR, so that GRB enhancement of inhibition, to extent it occurred, was confined to the first few meters of the water column. Thus, most of the water column was not affected by the increased UVB.

3) The greatest relative enhancement in UVB due to the GRB was in the winter, when incident UV and productivity are typically low. Generally, we were unable to assess the effects of winter time UV enhancement using the ocean color approach described here because of the lack of winter observations. However there could be some localized effects; this could be examined using existing field data and BWFs (e.g. Weddell-Scotia Confluence, Neale et al., 1998).

4) As explained in Neale et al. (2014), the weights for the shortest UV wavelengths most affected by GRB were probably overestimated for the earlier determined BWFs that were incorporated in the TUV radiative transfer model used by Thomas et al. (2015). The weights in question are at the short wavelength limit of these BWF estimates and Neale et al. (2014) found that this range is subject to artifacts of the estimation method. The *Synechococcus* and *Prochlorococcus* BWFs used here were estimated using a revised approach with an extended wavelength range which produced more robust estimates of weighting coefficients in the range most affected by GRBs.

We are confident in our estimates of GRB impacts on central ocean productivity since they are based on responses of the dominant primary producers in this system, which were analyzed to minimize uncertainty in the wavelength range most affected by ozone. The studied cultures are just a small sample of the global diversity of picophytoplankton in the temperature and tropical ocean (Scanlan et al., 2009), but the similarity of the results between strains suggest a general applicability of our results to these areas. Moreover, we don't expect that the broad conclusions drawn from the analysis would be that different using previously defined BWFs and response models.  For example, Neale et al. (2014) reported that a change in ozone column as might occur with a GRB from 300 to 200 DU under the open ocean conditions would result in a 0.1 to 0.3% inhibition effect on the productivity of *Synechococcus* or *Prochlorococcus* integrated to 0.1% LD, comparable to the average inhibition effect shown in Figs. 9 (a-c).  Using the model formulation and 5 literature BWFs described by Cullen et al. (2012) [two of which were also used by Thomas et al. (2015)], the predicted inhibition effect under the same conditions ranged from 0.3 to 0.9% [data not shown, used spreadsheet in Cullen et al. (2012) supplemental materials].  Both the $E_{max}$ model used here and an earlier version BWF/P-E model ($E$ model) used by Cullen et al. (2012) are irradiance based models, i.e. assume that inhibition is in equilibrium with exposure over the time scale of interest and thus time independent (Neale, 2000).  Where such models are applicable (i.e. most of the ocean, Neale and Kieber, 2000), the predicted inhibition effect of a GRB on integrated productivity is unlikely to be more than a few per cent.  A notable exception may be polar phytoplankton, brief consideration to their case will be given later on in the Discussion.

On the other hand, the whole-system approach taken in this analysis has revealed GRB effects that were not apparent in the Thomas et al. (2015) analysis.   For integrated water column production (0.1% penetration of $E'_{PAR}$), the effect of decreased atmospheric transparency in the UVA and short-wavelength visible due to increased $NO_2$ column had as much, or more, effect than the increase in UV due lower $O_3$ column. Partly this was due to lower availability of blue to blue-green light underwater, a spectral range that is particularly important to marine phytoplankton given their pigments absorb strongly in the 400-450 nm range. This decreased the amount of photosynthetically utilizable radiation (PUR) and productivity deeper in the water column. In contrast, there was little effect of the increased PAR due to less ozone absorbance in the Chappuis bands, as this occurs at longer wavelengths [500-700 nm, (Shaw, 1979)] for which there is neither high pigment absorbance (Larkum, 2003) nor high transparency of seawater (Pope and Fry, 1997). Since pigments in almost all marine phytoplankton absorb most strongly in the blue to blue-green band, this aspect of our results is not strongly tied to the choice of model species.

An additional effect of increased $NO_2$ is a reduction of UVA. Since UVA is particularly effective at causing inhibition in *Synechococcus* and *Prochlorococcus*, the reduction in UVA somewhat counterbalances the enhanced UVB, moderating the overall inhibition effect. In the time immediately after the event, this resulted in essentially no net effect of the GRB on integrated production, at least for the response based on the *Prochlorococcus* weighting function.

Some model limitations should be kept in mind when assessing the significance of these findings. Like most other models that estimate productivity based on ocean color, our integrated productivity model weights all depths equally even though the Chl profile frequently has a deep (below MLD) maximum in stratified regions (cf. Carr et al., 2006). However this deep biomass contributes little to integrated productivity and global comparison of productivity estimates by 24 models did not find a consistent difference between those with uniform vs. non-uniform Chl profiles (Carr et al., 2006). The expected effect of adding more weight to deeper depths, in any case, would be a (probably small) increase in the GRB effect due to enhancing the contribution of reduced PUR, which already dominates the overall effect on productivity. Also, like Thomas et al. (2015), our model predicts daily effects and we do not assess the cumulative effect of post-GRB conditions on phytoplankton biomass. Such an assessment might be possible by integrating our approach into models that include all sources and sinks of carbon in the marine food web, such as those discussed by Sailley et al. (2013).

While direct effects on photosynthesis seem to be minimal for central ocean phytoplankton, there are other GRB effects that could be important. The increase in DNA damage is very important (Thomas et al. 2015), and this could have implications for the growth of phytoplankton. Phytoplankton in the near surface zone may incur sufficient DNA damage to stop growth despite being able to continue photosynthesis (in the short term). The overall significance of this effect in the ocean has yet to be determined. Also, this analysis concentrated on the responses of temperate/tropical phytoplankton. The responses of polar phytoplankton are considerably different from phytoplankton in warmer waters due, in part, to slower rates of recovery after UV exposure introducing a time-dependence in their response (Fritz et al., 2008; Smyth et al., 2012). Polar phytoplankton are already episodically exposed to higher UVB due to the springtime "ozone hole". Integrated water column effect of springtime ozone depletion productivity is thought to be typically less than a few percent, and no more than 12% (Smith and Cullen, 1995). However, effects could be much higher if such depletion continued into the summertime period.  It should be possible to assess such local impacts using existing atmospheric modeling and BWF/P-Es appropriate to the Antarctic atmosphere and Southern Ocean near Antarctica.

Acknowledgments

Dirk Aurin, NASA for generously sharing his Matlab script for processing the QAA v6 algorithms. Cedric Fichot for assistance with attenuation coefficients. This work has been supported by the National Aeronautics and Space Administration under grant numbers NNX09AM85G and NNX14AK22G, through the Astrobiology: Exobiology and Evolutionary Biology Program. Computational time was provided by the High Performance Computing Environment (HiPACE) at Washburn University; thanks to Steve Black for assistance with computing resources.

Author Disclosure Statement – No competing financial interests exist

**Figure 1.** Variation of $T_{PIR}/T_{PUR}$ for the global ocean and selected regions. Annual average $T_{PIR}/T_{PUR}$ ratios estimated for all locations in the global ocean contained in the monthly $K_d$ climatology (see methods), expressed as zonal averages (±SD) over 10° latitude bins, vs. latitude (Global – red line). Shown for comparison are the same averages vs. latitude for the selected mid-Pacific (blue line) and East-Pacific (green line) longitude ranges.

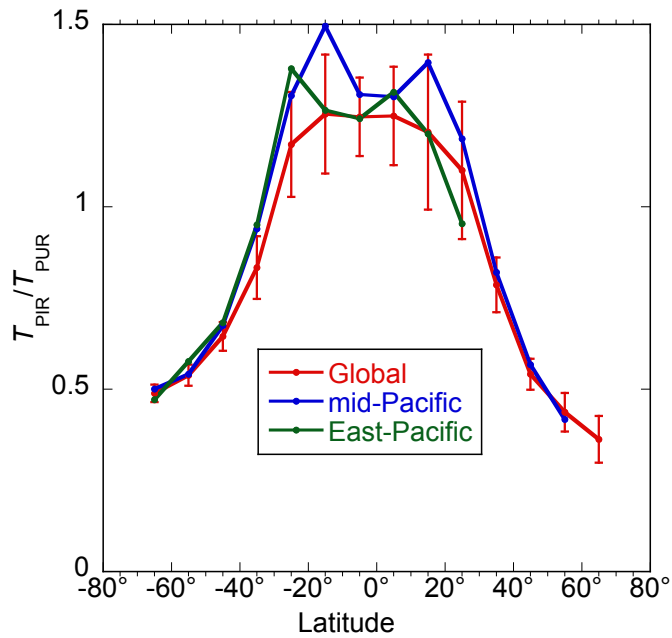

Figure 2. Latitude – time contour plots of the transparency (m) of inhibiting irradiance in the ocean surface layer ($T_{PIR}$) for two longitude regions in the Pacific for a five year period after a GRB event. The estimates used the biological weighting functions for inhibition of photosynthesis in two strains of *Synechococcus* (Syn1 and Syn2) and *Prochlorococcus* (Pro). Note that each panel has its own scale.

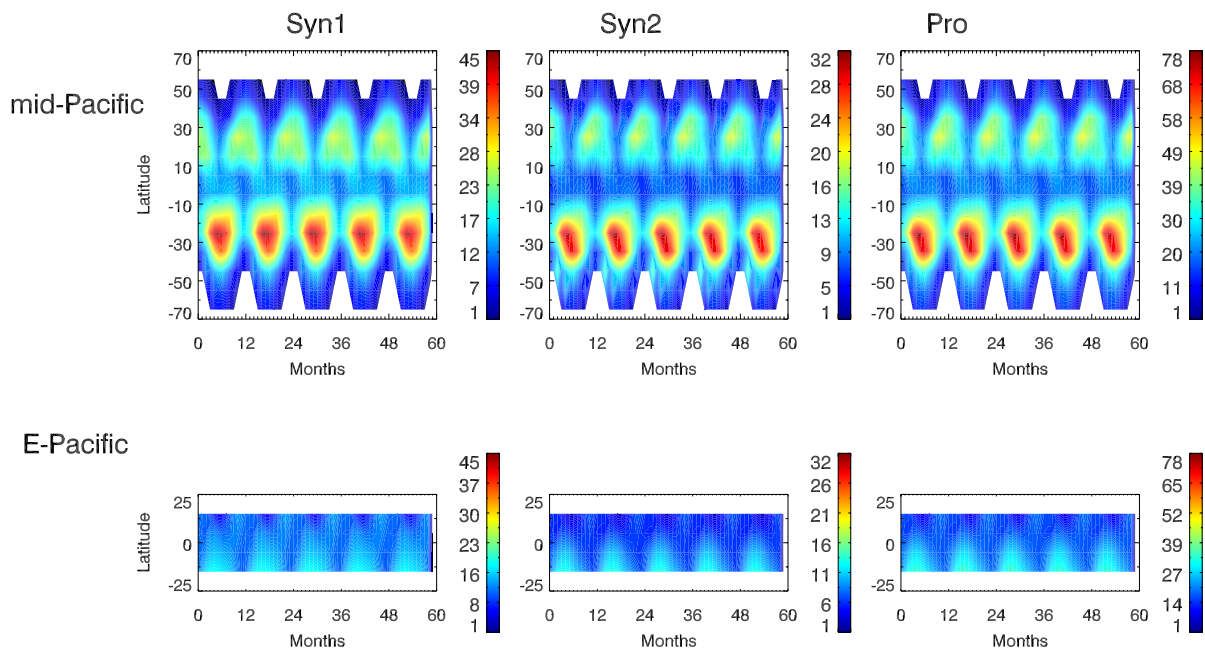

Figure 3. Latitude – time contour plots for the transparency (m) of photosynthetically utilizable radiation (PUR) in the ocean surface layer ($T_{\text{PUR}}$) for two longitude regions in the Pacific for a five year period after a GRB event. The estimates used the biological weighting functions for inhibition of photosynthesis in two strains of *Synechococcus* (Syn1 and Syn2) and *Prochlorococcus* (Pro). Note that each panel has its own scale.

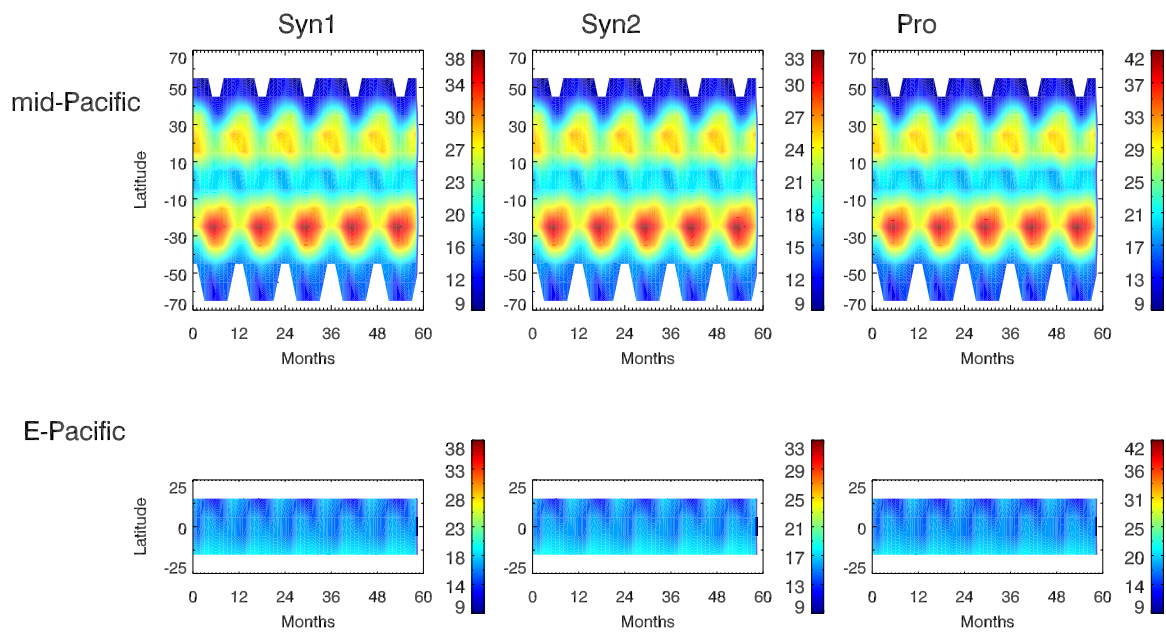

Figure 4  Average transparency (1 year running mean) calculated using model output at 25°S for photoinhibiting radiation ($T_{PIR}$) and photosynthetically utilizable radiation ($T_{PUR}$) over a five-year period following a Gamma Ray Burst.  Shown are the $T_{PIR}$ and $T_{PUR}$ means for *Synechococcus* (Syn1) for the mid-Pacific longitude band, similar patterns of variation were exhibited for transparencies calculated using the parameters for *Synechococcus* Syn2 and *Prochlorococcus*.

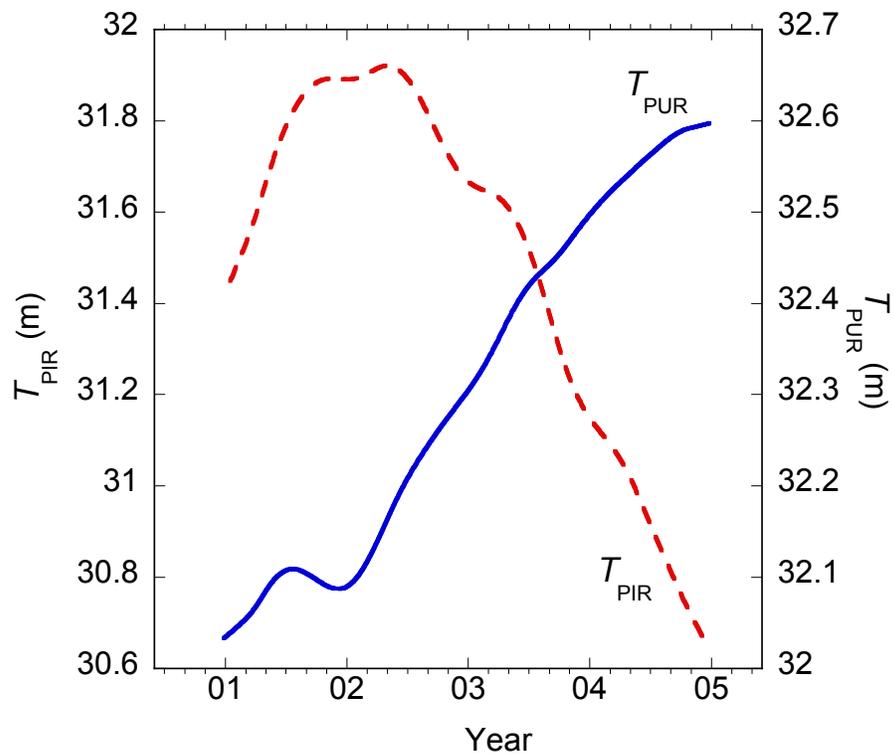

Figure 5. Effect of a GRB on phytoplankton productivity. Productivity, $P^B(z)$, integrated to the 0.1% light depth for the GRB case was normalized to productivity for "normal" conditions in the absence of a GRB. The GRB reduces productivity due to both more inhibition and less photosynthetically utilizable radiation. Results for two regions and three BWF types labeled as in Fig. 1.

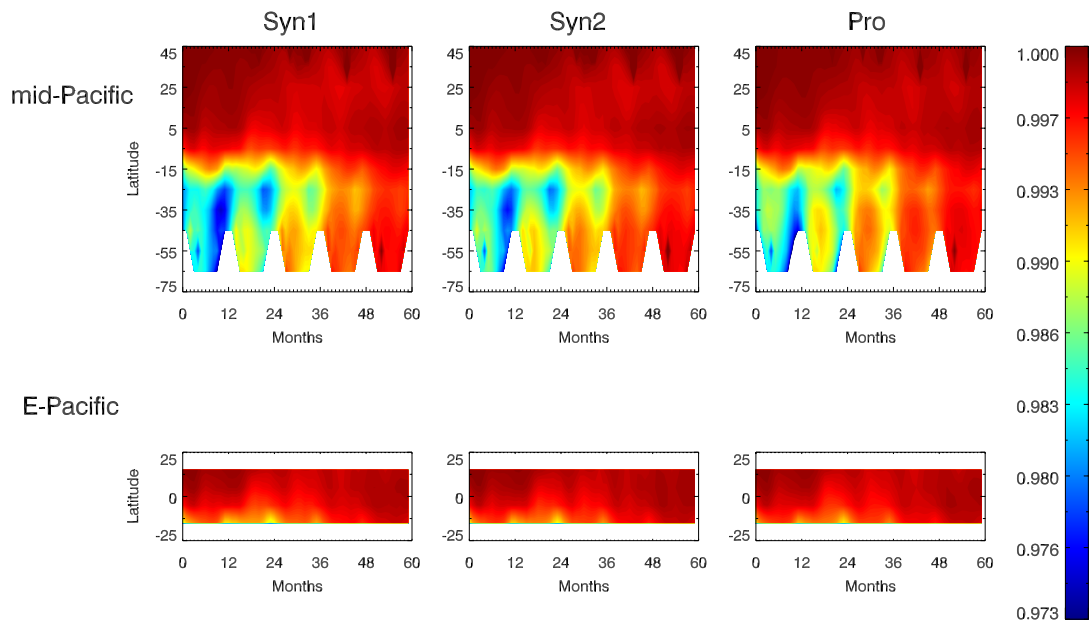

Figure 6. Effect of a GRB on potential phytoplankton productivity (in the absence of inhibition).   Potential productivity, $P_{pot}^{B}$ $(z)$, integrated to the 0.1% light depth for the GRB case was normalized to productivity for "normal" conditions in the absence of a GRB.  The GRB reduces potential productivity due to less photosynthetically utilizable radiation.  Results for two regions and three BWF types labeled as in Fig. 1.

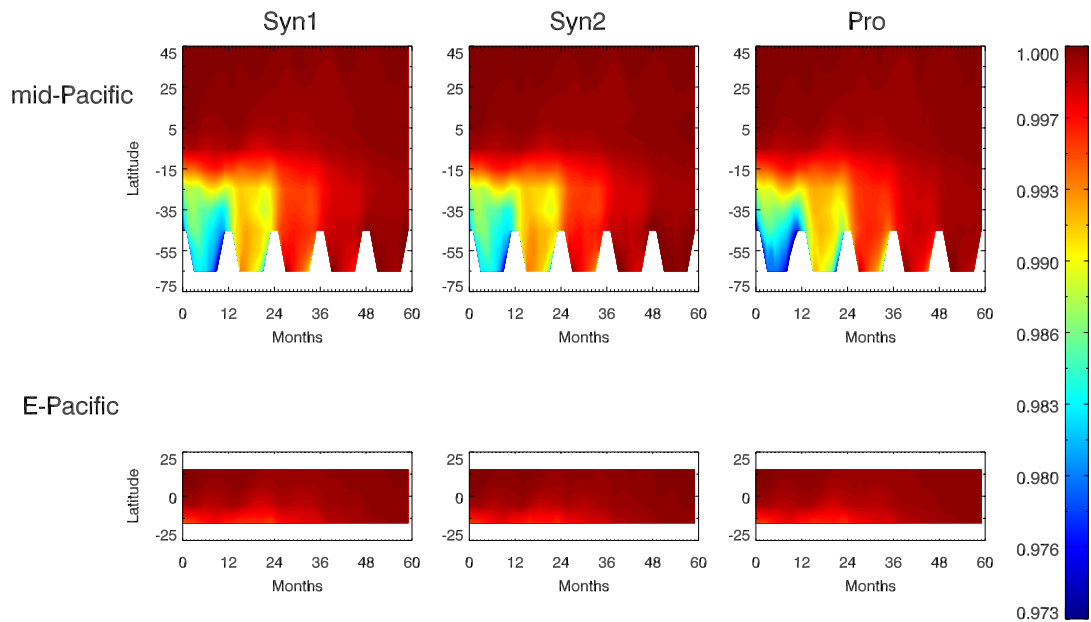

Figure 7.  Effect of a GRB on phytoplankton productivity integrated to the mixed layer depth.    Productivity, $P^B(z)$, integrated to the mixed layer depth for the GRB case was normalized to productivity for "normal" conditions in the absence of a GRB. The GRB reduces productivity due to both more inhibition and less photosynthetically utilizable radiation.  Results for two regions and three BWF types labeled as in Fig. 1.

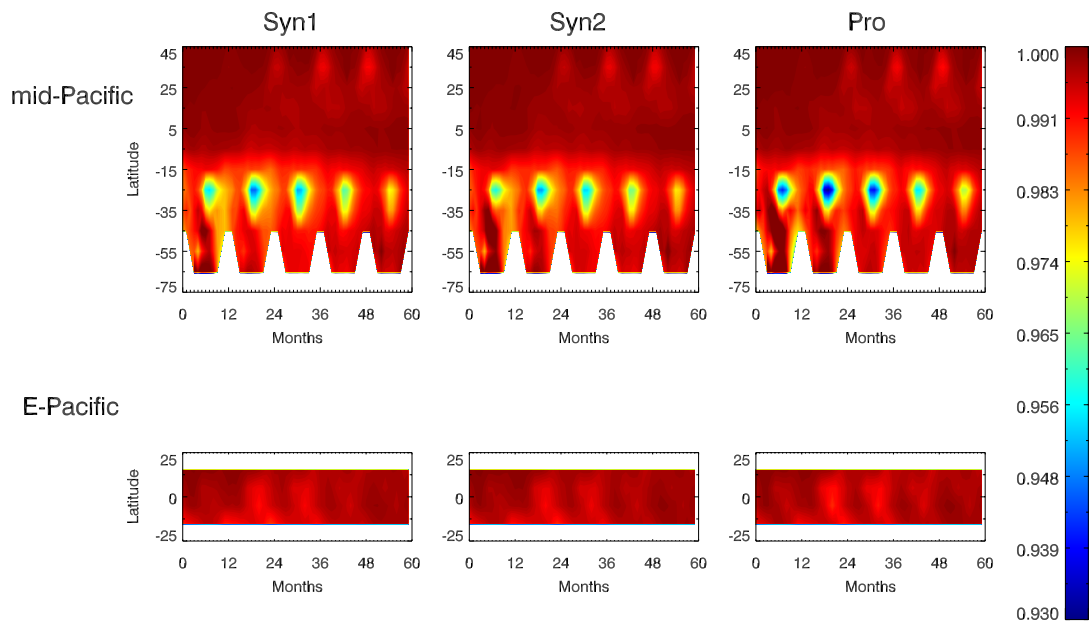

Figure 8. Effect of a GRB on potential phytoplankton productivity integrated to the mixed layer depth. Potential productivity, $P_{pot}^{B}$ *(z)*, integrated to the mixed layer depth for the GRB case was normalized to productivity for "normal" conditions in the absence of a GRB. The GRB reduces potential productivity due to less photosynthetically utilizable radiation. Results for two regions and three BWF types labeled as in Fig. 1.

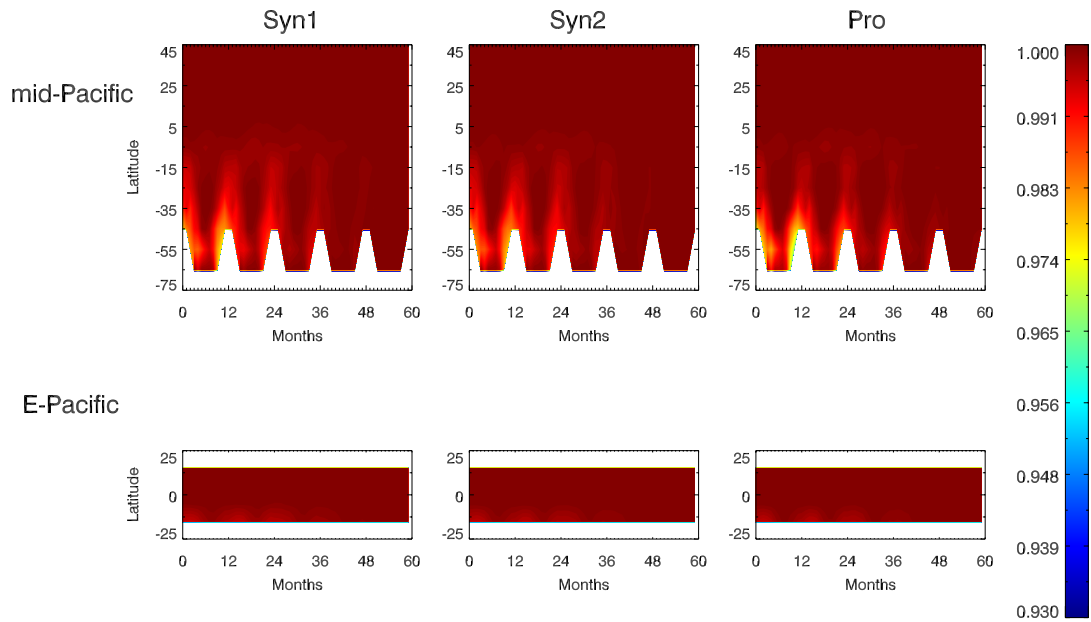

Figure 9. Weighted average productivity (monthly means) following a GRB by calendar year following the hypothetical June 2000 event date. Shown is the ratio of productivities calculated for GRB and no-GRB cases averaged over the mid-Pacific longitude S. Hemisphere. Productivity at each location was weighted by product of daylength and Chl to be approximately proportional to the contribution to total hemispheric productivity at each time (see details in Materials and Methods). Productivity was averaged both including ($<P>_Z$- blue) and excluding ($<P>_{pot-Z}$- red) inhibition due to UV and PAR. The difference between the two averages is considered the average effect of inhibition. The estimates used the biological weighting functions for inhibition of photosynthesis in two strains of *Synechococcus* (Syn1 and Syn2) and *Prochlorococcus* (Pro), and were performed for productivities integrated to the 0.1% light depth for $E'_{PAR}$ (A-C) and the mixed layer depth (D-F.)

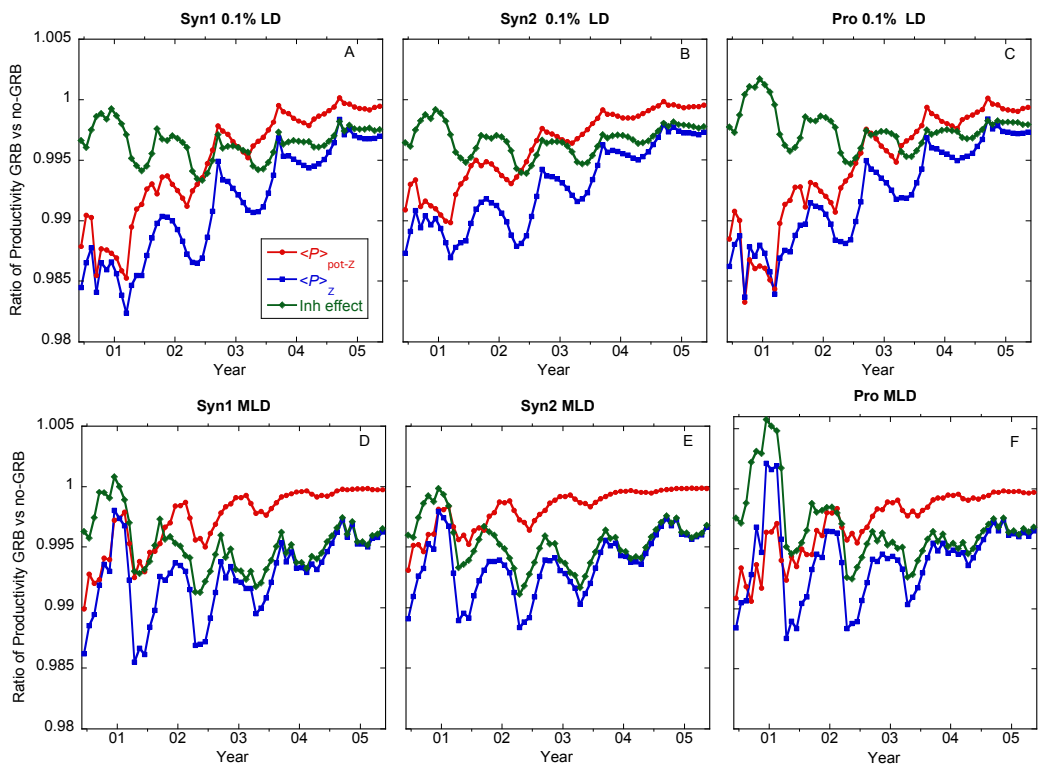